\documentclass[aps,prd,onecolumn,showpacs,showkeys,nofootinbib,notitlepage]{revtex4-1}
\usepackage{epsfig}
\usepackage{amsmath}
\usepackage{amsfonts}
\usepackage{amssymb}
\usepackage{graphicx}
\usepackage{colordvi}
\usepackage{ float}
\usepackage{colortbl}
\usepackage{bm}
\usepackage{latexsym}
\usepackage{psfrag}
\usepackage{amsthm}
\usepackage{color}
\usepackage{braket}
\usepackage{indent first}
\usepackage[normalem]{ulem}
\usepackage{fancyhdr}
\usepackage{hyperref}
 \usepackage{environ}
\usepackage{mathtools}
\usepackage{float}

\renewcommand\[{\begin{equation}}
\renewcommand\]{\end{equation}}
\newcommand{\be}{\begin{equation}}
\newcommand{\ee}{\end{equation}}
\newcommand{\bea}{\begin{eqnarray}}
\newcommand{\eea}{\end{eqnarray}}

\newcommand{\g}{\gamma}

\newcommand{\ba}{\begin{eqnarray}}
\newcommand{\ea}{\end{eqnarray}}

\begin{document}

\title{Modified dispersion relations and a potential explanation of the 
EDGES anomaly}


\author{Saurya Das} 
\email{saurya.das@uleth.ca}
\affiliation{Theoretical Physics Group and Quantum Alberta, Department of Physics and Astronomy,
University of Lethbridge,
4401 University Drive, Lethbridge,
Alberta, T1K 3M4, Canada}

\author{Mitja Fridman}
\email{fridmanm@uleth.ca}
\affiliation{Theoretical Physics Group and Quantum Alberta, Department of Physics and Astronomy,
University of Lethbridge,
4401 University Drive, Lethbridge,
Alberta, T1K 3M4, Canada}

\author{Gaetano Lambiase} \email{lambiase@sa.infn.it}
\affiliation{Dipartimento di Fisica E.R: Caianiello, Universita di Salerno, Via Giovanni Paolo II, 132 - 84084 Fisciano, Salerno, Italy \& INFN - Gruppo Collegato di Salerno, Italy.}

\author{Antonio Stabile} \email{anstabile@gmail.com}
\affiliation{Dipartimento di Fisica E.R: Caianiello, Universita di Salerno, Via Giovanni Paolo II, 132 - 84084 Fisciano, Salerno, Italy \& INFN - Gruppo Collegato di Salerno, Italy.}

\author{Elias C. Vagenas} \email{elias.vagenas@ku.edu.kw}
\affiliation{Theoretical Physics Group, Department of Physics, Kuwait University, P.O. Box 5969, Safat 13060, Kuwait}

%
%
%
%
%
\begin{abstract}
\par\noindent
The \emph{Experiment to Detect the Global Epoch of Reionisation Signature} (EDGES) collaboration 
has recently reported an important result related to the absorption signal in the Cosmic Microwave Background radiation spectrum.
This signal corresponds to the red-shifted 21-cm line at $z \simeq 17.2$, whose amplitude is about twice the expected value.
This represents a deviation of approximately $3.8\sigma$ from the  predictions of the standard model of cosmology, i.e. the
$\Lambda$CDM model. This opens a window for testing new physics beyond both the standard model of particle physics and the $\Lambda$CDM model. 
%
In this work, we explore the possibility of explaining the EDGES anomaly in terms of modified dispersion relations. 
The latter are typically induced in unified theories and theories of quantum gravity, such as String/M-theories and Loop Quantum Gravity. 
%
These modified dispersion relations affect the density of states per unit volume and thus the thermal  spectrum of the Cosmic Microwave Background photons. The temperature of the 21-cm brightness temperature is modified accordingly giving a potential explanation
of the EDGES anomaly.
%

\end{abstract}
\date{\today}
\maketitle

%
%
%
%
%
\section{Introduction}
%
%
%
%
%
\par\noindent
Predictions of General Relativity (GR) have been tested with high accuracy ranging from the solar system to the cosmological scales. 
Despite this success, GR is an incomplete theory at short distance and time scales 
(for example, near black holes and cosmological singularities), and perhaps at large distances as well, where dark components and/or modifications of GR are invoked to explain the accelerated phase of the present Universe.
It is expected that the inconsistencies at small scales can be resolved within the framework of quantum gravity (QG), which incorporates
the principles of GR and quantum theory, and provides a description of the microstructure of space-time at the Planck scale.

Among the various attempts towards formulating a theory of QG, String/M-theory and Loop Quantum Gravity (LQG) remain as important candidates.
A consequence of these theories is that space-time appears non-commutative (NC) at the fundamental level
\cite{Banks:1996vh,Seiberg:1999vs,Sheikh-Jabbari:1999cvv, Maldacena:2000vw,Chamseddine:2001hk}, and in some situations, may
also exhibit a varying speed of light \cite{Hashimoto:2000ys,Alexander:2001dr}. 
%
This gives rise to non-local field theories and 
a modification of the dispersion relation of the quantum fields in a NC space-time.
For example, one of the consequences of String Theory (as well as of M-Theory) is space-time non-commutativity
\cite{Banks:1996vh}, with the 
latter leading to modified dispersion relations \cite{Kempf:2000ac}.
Related to this is the fact that owing to quantum fluctuations, the usual canonical commutation relation also gets modified $[x,p]=i\hbar(1+\beta p^{2})$
\cite{Capozziello:1999wx,Chu:2000ww,Easther:2001fi,Kempf:2001fa,luciano,luciano1} (see also Refs. \cite{Brandenberger:2000wr,Martin:2000xs, Das:2008kaa, Ali:2011fa}).
It must be pointed out however, that there are several other approaches to QG that also predict the existence of a minimum measurable length, which in turn 
represents a natural cutoff and induces a departure from the relativistic dispersion relation.
These approaches include space-time foam models \cite{Amelino-Camelia:1997ieq,Amelino-Camelia:1999jfz, Amelino-Camelia:2000stu}, spin-network in LQG \cite{Gambini:1998it}, space-time discreteness \cite{tHooft:1996ziz},
spontaneous symmetry breaking of Lorentz invariance in string field
theory \cite{Kostelecky:1988zi} or in NC geometry \cite{Carroll:2001ws}, Horava's approach \cite{Horava:2009uw,Horava:2009if}, and
Doubly Special Relativity (DSR) \cite{Amelino-Camelia:2000stu,Amelino-Camelia:2000cpa, Amelino-Camelia:2010lsq}. In Ref. \cite{Magueijo:2002xx}, the authors proposed an extension of DSR to include curvature, also known as Doubly General Relativity, in such a way that the geometry of space-time does depend on the energy $E$ of the particle used to probe it (\emph{gravity's rainbow}) \cite{Amelino-Camelia:2000stu}.
%
The general form of the modified dispersion relation (MDR) reads \cite{Ali:2014aba}
\begin{equation}\label{disprela1}
{{E}^{2}}f{{\left(E/{{E}_{P}} \right)}^{2}}-{{p}^{2}}\,c^2\,g{{\left( E/{{E}_{P}} \right)}^{2}}=m^{2}c^4,
\end{equation}
where the (rainbow) functions $f(E/E_P)$ and $g(E/E_P)$ depend on the Planck energy $E_P=1.221\times10^{19}\,\text{GeV}$ (for details see for instance Refs. \cite{Liu:2007fk,Leiva:2008fd,Li:2008gs,Garattini:2011hy,Awad:2013nxa,Barrow:2013gia,Ali:2014xqa,Ali:2014cpa}).
Now, whenever $f,g \neq 1$, i.e., one deviates from the standard relativistic dispersion relation, as we shall show below, the Planck radiation spectrum changes as well. 
This in turn may be able to explain the anomaly, 
which the {\em Experiment to Detect the Global Epoch of Reionisation Signature} (EDGES) collaboration has recently reported \cite{Bowman:2018yin}. In the range $z=15-20$, the EDGES collaboration found an anomalous absorption profile, with a brightness temperature minimum 
at $z_{\rm E} \simeq 17.2$, which has a magnitude of about a factor of two greater than predicted by the $\Lambda$CDM model. It is this anomaly that we propose to explain using MDRs in this work. 
It turns out that the standard MDRs do not adequately explain the EDGES anomaly.
However, by imposing redshift dependent MDR parameters, or by imposing a non-trivial power dependence for the MDRs, we are able to provide a viable explanation for the EDGES anomaly. A non-trivial power dependence of a MDR is also discussed in Ref. \cite{Arzano:2016twc}.

The rest of the paper is organized as follows. 
In the next section, we briefly review some of the important special cases of the above MDR.
Following this, in Section \ref{experiments}, 
we estimate the parameters in the models that we consider from the results of the EDGES experiment. 
Finally, we summarize our results and conclude in Section \ref{conclusions}.
%
%
%
%
%
\section{MDR and modification of thermal spectrum}
\label{mdr}
%
%
%
%
%
%
%
\par\noindent
As stated in the Introduction, MDR is predicted by various theories of QG, and has the most general form of Eq. (\ref{disprela1}). The rainbow functions can in the most general case be expressed in a power series expansion (MacLaurin series) as $f\left( E/{{E}_{P}} \right)=\sum_{n=0}^\infty\frac{f^{(n)}(0)}{n!}\left(E/E_P\right)^n$ and $g\left( E/{{E}_P} \right)=\sum_{n=0}^\infty\frac{g^{(n)}(0)}{n!}\left(E/E_P\right)^n$, where constraints $f(0)=1$ and $g(0)=1$ must be imposed to obtain the standard relativistic dispersion relation at low energies. Here we consider some of the interesting special cases.

\begin{itemize}
\item Case 1: 
$f\left(E/{{E}_{P}} \right)=1$, $g\left( E/{{E}_{P}} \right)=\sqrt{1-\eta\, (E/E_P)^\omega}$, 
which is one of the most studied in literature. 
Here $\eta$ is a parameter which signifies 
the effective scale of the modification, and $\omega$ is the order of the modification. 
A complete theory of QG should fix both of them. However, in this work we study the modifications for different values 
of $\eta$ and $\omega$, and in particular, 
we consider three special cases.
The first case is compatible with LQG and NC space-time \cite{Amelino-Camelia:1996bln,Amelino-Camelia:2008aez}, while the next two are compatible with the linear and quadratic Generalized Uncertainty Principle (GUP) respectively \cite{Majhi:2013koa,Das:2021lrb}:
\begin{itemize}
\item[i)] $\omega=1$ and $\eta>0$ $\Longrightarrow$ $f\left(E/{{E}_{P}} \right)=1$, $g\left( E/{{E}_{P}} \right)=\sqrt{1-\eta\, E/E_P}~,$
\item[ii)] $\omega=1$ and $\eta=\mp2\alpha_0$ $\Longrightarrow$ $f\left(E/{{E}_{P}} \right)=1$, $g\left( E/{{E}_{P}} \right)=\sqrt{1\pm2\alpha_0\, E/E_P}~,$
    \item[iii)] $\omega=2$ and $\eta=2\beta_0$ $\Longrightarrow$ $f\left(E/{{E}_{P}} \right)=1$, $g\left( E/{{E}_{P}} \right)=\sqrt{1-2\beta_0\, (E/E_P)^2}~,$
\end{itemize}
where restrictions on $\alpha_0$ from Ref. \cite{Majhi:2013koa} have been relaxed to include both positive and negative values.
In general, $f\left( E/{{E}_{P}} \right)\neq1$  and, specifically, in the presence of a strong gravitational field
$f\left( E/{{E}_{P}} \right)=1/\sqrt{-g_{00}}$, where $g_{00}$ is the $00$ component of the metric \cite{Majhi:2013koa,Das:2021lrb}. 
However, in the dark ages, most of the hydrogen gas was in a very weak field, and, therefore, we can set $f\left( E/{{E}_{P}} \right)=1$, as far as space-time curvature corrections to the MDR are concerned. 

\item Case 2: $f\left( E/{{E}_{P}} \right)=\frac{{{e}^{\alpha E/E_P}}-1}{\alpha E/{{E}_{P}}},$ $g\left( E/{{E}_{P}} \right)=1$, proposed for explaining the spectra from GRBs at cosmological distances \cite{Amelino-Camelia:1997ieq}.

\item Case 3: $f\left(E/{{E}_{P}}\right)=1$, $g\left(E/{{E}_{P}}\right)=[1+(\lambda E)^\gamma]^\delta =[1+\lambda'\, (E/E_P)^\gamma]^\delta$, with $\lambda'=(\lambda E_P)^\gamma$. The case $\delta=1$ has been proposed for models in which a varying speed of light occurs \cite{Alexander:2001dr}.
    The case $\gamma=\delta=1$ has been proposed in Refs. \cite{Amelino-Camelia:2000stu,Alexander:2001ck}.
    For $\gamma=1$, $\delta=1/2$, and $\lambda=-\eta$, we recover Case 1. GUP provides another case for this form of $g\left(E/{{E}_{P}}\right)$ \cite{Das:2021lrb}:
    \begin{itemize}
        \item[i)] $\delta=1$, $\gamma=1$ and $\lambda'=\pm\alpha_0$ $\Longrightarrow$ $f\left(E/{{E}_{P}} \right)=1$, $g\left(E/{{E}_{P}}\right)=1\pm\alpha_0\, E/E_P$~.
    \end{itemize}

\end{itemize}
%
%
Here $\eta, \alpha_0, \beta_0, \alpha, \lambda'$ are dimensionless parameters, with $\alpha_0$ and $\beta_0$ to be the linear and quadratic GUP parameters, respectively. It is often assumed that $\eta, \alpha_0, \beta_0, \alpha,\lambda'\sim {\cal O}(1)$,
so that the modifications of the dispersion relations are non-negligible at Planck scales.
However, one may relax such a restriction and investigate signals of new physics at a new intermediate scale
$\lambda_{new}\sim {\eta}{\ell_{P}} \sim {\alpha_0}{\ell_{P}} \sim {\sqrt{\beta_0}}{\ell_{P}} \sim {\alpha}{\ell_{P}} \sim {\lambda'}{\ell_{P}}$. Such a length (energy) scale $\lambda_{new}$
cannot exceed the well-tested electroweak length scale, $\lambda_{EW}\sim {10^{17}}{\ell_{P}}$, so the consequent upper bound $\eta\sim \alpha_0 \sim \sqrt{\beta_0}\sim \alpha \sim \lambda' \leq 10^{17}$.

The MDR given by Eq. (\ref{disprela1}), for the case of photons reads
\be\label{disp}
E^2 - p^2 c^2 F^2 =0\,, \quad \text{with} \,\, F=\frac{g}{f}\,,
\ee
so that, following Refs. \cite{Alexander:2001ck,Alexander:2001dr}, one may derive the modified thermal spectrum $\rho_{MDR}$. 
\par\noindent
The density of states per volume for photons (which have 2 polarization states) is written as 
\begin{eqnarray}
\Omega(p)=\frac{p^2}{\pi^2\hbar^3}~.
\end{eqnarray}
By considering the MDR in Eq. (\ref{disp}) and using $\Omega(E)\,\mathrm{d}E=\Omega(p)\,\mathrm{d}p$, we obtain the following density of states 
%
\begin{eqnarray}
\Omega(E)=\frac{E^2}{\pi^2\hbar^3\hat{c}^2\Tilde{c}}~,
\end{eqnarray}
where the two `speeds' in the above equation turn out to be
\begin{eqnarray}
\hat{c}=\frac{E}{p}=cF\,\,\,\,\,\,\mathrm{and}\,\,\,\,\,\,\Tilde{c}=\frac{\mathrm{d}E}{\mathrm{d}p}=\frac{cF}{1-\frac{F'E}{F}}~,
\end{eqnarray}
where $F'=\mathrm{d}F/\mathrm{d}E$. Therefore, we can write the modified density of states as
\begin{eqnarray}
\label{dsmdr}
\Omega(E)=\frac{E^2}{\pi^2\hbar^3c^3}\frac{1}{F^3}\left|1-\frac{F'E}{F}\right|~.
\end{eqnarray}
The modified thermal spectrum is then obtained using\footnote{We use this definition to obtain the standard thermal spectrum, as can be found in \cite{caniou,sharkov}. This differs from the definition used in \cite{Alexander:2001ck} by an unimportant factor of $2\pi\hbar$, which has no effect on our calculations or results.} $\rho_{MDR}(T,E)=2\pi\hbar E\,n(E)\,\Omega(E)$, where $n(E)=\frac{1}{e^{\beta_T E}-1}$ is the Bose-Einstein distribution, $\beta_T=\frac{1}{k_BT}$ is the inverse temperature and $k_B$ is the Boltzmann constant. The modified thermal spectrum then reads as
\begin{equation}\label{rhoNC}
\rho_{MDR}(E,T) = \rho(E,T) \frac{1}{F^3}{\left| 1-\frac{F'E}{F}\right|}\equiv \rho(E,T)R~,
\end{equation}
where
\begin{equation}\label{radiation}
\rho(E,T)=\frac{2}{\pi\hbar ^2 c^3}\frac{E^3}{e^{\beta_T E}-1}
\end{equation}
is the standard thermal distribution of photons and $R$ is the correction factor, formally defined in the following section. Note that the standard result from Eq. (\ref{radiation}) is obtained from Eq. (\ref{rhoNC}) when the MDR parameters vanish, i.e., $\eta, \alpha_0, \beta_0, \alpha, \lambda'\longrightarrow0$.
%
%
%
\section{Experimental bounds}
\label{experiments}
%
%
%
%
\par\noindent
In this section we study the effects of the modified thermal spectrum given by Eq. (\ref{rhoNC}), induced by the MDR given in Eq. (\ref{disp}), on the 21-cm cosmology. Details of 21-cm cosmology are given in Appendix \ref{appendix}.
This is related to the history of the universe, and represents a new framework for probing fundamental physics \cite{Barkana:2018lgd} (for other models see Refs. \cite{Furlanetto:2006jb,Fornengo:2011cn,Fixsen:2009xn,Pritchard:2011xb,Lopez-Honorez:2016sur,AristizabalSierra:2018emu,Hill:2018lfx,Pospelov:2018kdh, Moroi:2018vci, Lambiase:2018lhs}). In particular, we focus on the recent release of the EDGES 
collaboration \cite{Bowman:2018yin} (see also Ref. \cite{Chianese:2018luo}).

EDGES High and Low band antennas probe the frequency ranges 90-200 MHz and 50-100 MHz, respectively, overall measuring the 21-cm signal within the redshift range $z\in 6 - 27$,
corresponding to an age of the Universe $t_U\in (100 {\rm Myr} - 1 {\rm Gyr})$, i.e., the {\em dark ages}.
This includes the epochs of reionization and cosmic dawn,
in which the 
first astrophysical sources form. 
%
At $z_{\rm E} \simeq 17.2$, the observed
magnitude of the absorption line\footnote{The EDGES collaboration found an absorption profile approximately in the range $z = 15-20$, with the minimum at the redshift $z_E = 17.2$.} is
about a factor of two greater than the one predicted by the $\Lambda$CDM model.
At the redshift of the minimum of the 21-cm line, i.e., $z_{\rm E} \simeq 17.2$, and frequency of CMB radiation, i.e., 
$\nu_{21}(z_{\rm E})\simeq 78\,{\rm MHz}$, one has a 21-cm brightness temperature $T_{21}(z_{\rm E})=-0.5_{-0.5}^{+0.2}\,{\rm K}$ ($99\%\,{\rm C.L.}$, including estimates of systematic uncertainties).
Since at $z=z_E$ one has $(1+\delta_{\rm b})\,x_{H_I}(z_{\rm E}) \simeq 1$, Eq. (\ref{T21}) implies $T_{\g}(z_{\rm E})/T_S(z_{\rm E}) =15^{+15}_{-5.5}$ \cite{Chianese:2018luo,Barkana:2018lgd}. 
Moreover, in the context of the $\Lambda$CDM model, one also gets
 \[
T_{\g}(z_{\rm E}) = T_{CMB}(z_{\rm E}) = T_{CMB,0}\,(1+z_{\rm E}) \simeq 50\,{\rm K}
\]
and
\[
T_{\rm gas}(z_{\rm E}) \simeq T_{CMB}(z_{\rm dec}^{\rm gas})\,\left(\frac{1+z_{\rm E}}{1+z_{\rm dec}^{\rm gas}}\right)^2 \simeq 6\,{\rm K}~,
\]
where $z_{\rm dec}^{\rm gas}\simeq 150$ and $T_{\rm dec}^{\rm gas} \simeq 410\,{\rm K}$ are
the  redshift and the temperature at the time when the gas and radiation decouple. Using (\ref{T21}), one infers
$T_{21}(z_{\rm E}) \gtrsim  -0.2\,{\rm K}$. Notice that the minimum is saturated for $T_{S}(z_{\rm E}) =
T_{\rm gas}(z_{\rm E})$, which corresponds to $T_{\g}(z_{\rm E})/T_{\rm gas}(z_{\rm E}) \simeq 8$.
As a consequence of these results, one finds that the best fit value for $T_{21}(z_E)$ is $\sim 2.5$ times lower than expected within the $\Lambda$CDM.

The 21-cm CMB photons  absorbed at $z_E$ fall clearly in the Rayleigh-Jeans tail
since $E_{21} \ll k_BT(z_{\rm E})$, where $E_{21}$ is the hyperfine transition energy of the hydrogen atom. The energy density of the photons, i.e., Eq. (\ref{radiation}), evaluated at $T=T_{CMB}$, reads
\be\label{rhoCMB}
\rho_{CMB}(E,z) =  \frac{2}{\pi\hbar^2 c^3}\,\frac{E^3}{e^{\beta_{T_{CMB}}\!(z)\,{E}}-1}~, 
\ee
where $\beta_{T_{CMB}}\!(z)=\frac{1}{k_BT_{CMB}(z)}$. Only photons with energy $E_{21}$ at $z\simeq z_{\rm E}$ could be absorbed by the neutral hydrogen producing
a 21-cm absorption global signal.
For explaining the EDGES results, we consider the $\rho_{MDR}$ given by Eq. (\ref{rhoNC}).
Therefore, we define the parameter $R$ to study the discrepancy from the $\Lambda$CDM model as
\be\label{R}
R \equiv \frac{\rho_{MDR}(E_{21},z_E)}{\rho_{CMB}(E_{21},z_E)}=\frac{1}{F^3}{\left| 1-\frac{F'E_{21}}{F}\right|}~, 
\ee
with $\rho_{MDR}$ and $\rho_{CMB}$ defined in Eqs. (\ref{rhoNC}) and (\ref{rhoCMB}), respectively. It may appear that 
such a modification may affect the optical depth $\tau_\nu$ (introduced in Appendix \ref{appendix}) and, therefore, the intensity and shape of the 21-cm line profile. However, as shown in Appendix \ref{appendix1}, such a modification does not affect $\tau_\nu$ in any way.
The experimental values from the EDGES experiment can then be explained by imposing
(see Ref. \cite{Chianese:2018luo} for details)

\be\label{Rexp}
R = 2.15^{+2.15}_{-0.8} \,.
\ee
Parameter $R$ in Eq. (\ref{R}) is then only a function of $F$, $F'$ and $E$, since everything else except the relevant correction cancels out. Since we can in general write the rainbow functions $f$ and $g$ as a power series in $E/E_P$, we can also write the function $F=g/f$ as a power series expansion
\begin{eqnarray}
F\left( E/{{E}_{P}} \right)=\frac{g\left( E/{{E}_{P}} \right)}{f\left( E/{{E}_{P}} \right)}=\sum_{n=0}^\infty\frac{F^{(n)}(0)}{n!}\, \left(E/E_P\right)^n~.
\end{eqnarray}
Note that 
$F(0)=1$, 
which corresponds to the 
standard $\Lambda$CDM result. 
The parameter $R$ from Eq. (\ref{R}) for such a general expression reads
\begin{eqnarray}
R=\frac{\left|1-\sum_{n=1}^\infty\frac{F^{(n)}(n-1)}{n!}\left({E}/{E_P}\right)^{n}\right|}{\left[\sum_{m=0}^\infty\frac{F^{(m)}}{m!}\left({E}/{E_P}\right)^{m}\right]^4}~.
\label{R2a}
\end{eqnarray}
Either Eq.(\ref{R}) or Eq.(\ref{R2a}) above can be used to 
estimate $R$ for the cases studied here,
compare with experimentally measured values and obtain bounds on the various parameters. 
%
%
%
%
\begin{itemize}

\item Case 1: $f\left( E/{{E}_{P}} \right)=1$, $g\left( E/{{E}_{P}} \right)=\sqrt{1-\eta\, (E/E_P)^\omega}~$. The ratio $R$ reads

\begin{eqnarray}
R=\frac{|1-\left(1-\frac{\omega}{2}\right)\eta\, (E/E_P)^\omega|}{(1-\eta\, (E/E_P)^\omega)^{5/2}}~
\end{eqnarray}
for arbitrary parameters $\eta$ and $\omega$. We take a look at the special cases:

\begin{itemize}
\item[i)] For $\omega=1$ and $\eta>0$, we have
 \begin{eqnarray}
 R=\frac{|1-\eta\, E/2E_{P}|}{(1-\eta\, E/E_P)^{5/2}}~.
 \label{R2}
 \end{eqnarray}

The ratio $R$ is plotted as a function of $\eta$ in Fig. \ref{REPL1}. 
To fit the EDGES experimental bounds, the parameter is fixed at
$\eta= 6.5_{-3.6}^{+4.0}\times10^{32}$.
 \begin{figure}[H]
 \centering
  \includegraphics[scale=0.8]{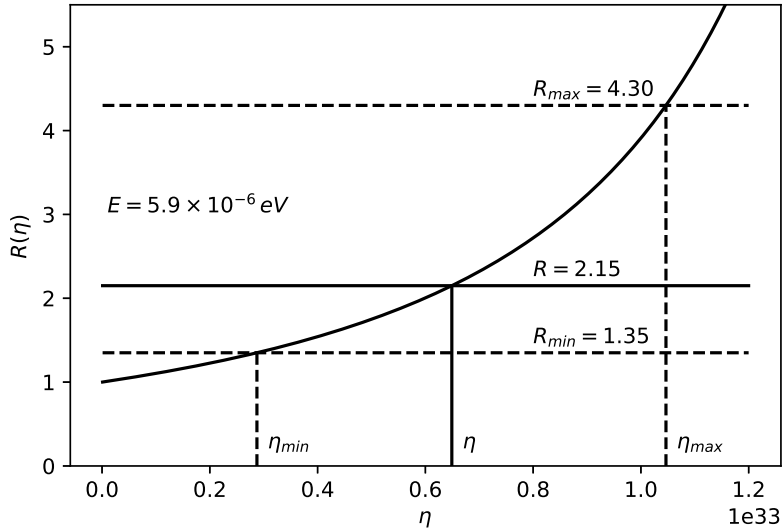}\\
  \caption{$R$ vs $\eta$ for fixed energy $E=E_{12}\simeq 5.9\times 10^{-6}$eV. }\label{REPL1}
\end{figure}
    \item[ii)] For $\omega=1$ and $\eta=\mp2\alpha_0$ we have
    \begin{eqnarray}
    R =\frac{|1\pm\alpha_0\, E/E_P|}{[1\pm2\alpha_0\, E/E_P]^{5/2}}~.
    \end{eqnarray}
    The ratio $R$ is plotted as a function of $\alpha_0$ for both branches in Fig. \ref{Ralpha}. However, only the branch with $\eta=+2\alpha_0$ can fix $\alpha_0$. To fit the EDGES experimental bounds, the parameter is fixed at $\alpha_0=3.2_{-1.8}^{+2.0}\times10^{32}$.
    \begin{figure}[H]
 \centering
  \includegraphics[scale=0.8]{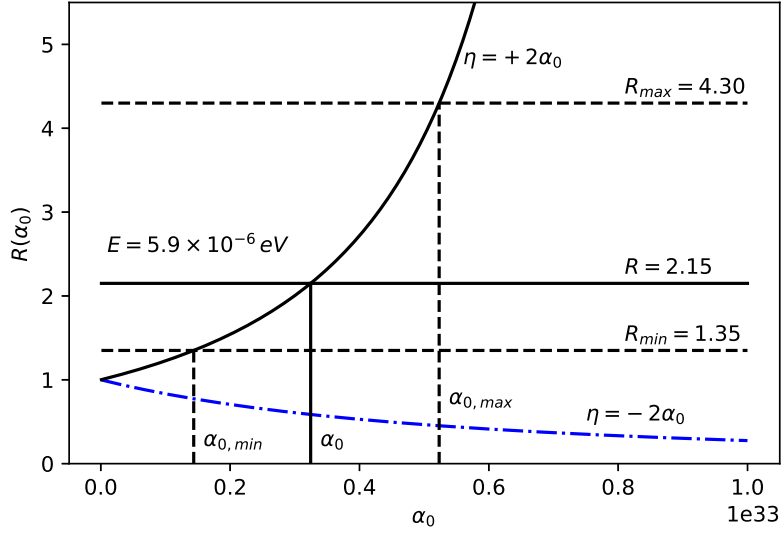}\\
  \caption{$R$ vs $\alpha_0$ for fixed energy $E=E_{12}\simeq 5.9\times 10^{-6}$eV. The $\eta=-2\alpha_0$ branch is presented in dash-dot blue and the $\eta=+2\alpha_0$ branch is presented in solid black.}\label{Ralpha}
\end{figure}
\item[iii)] For $\omega=2$ and $\eta=2\beta_0$ we have 
\begin{eqnarray}
R=\frac{1}{(1-2\beta_0\, (E/E_P)^2)^{5/2}}~.
\end{eqnarray}
The ratio $R$ is plotted as a function of $\beta_0$ in Fig. \ref{Rbeta}. 
To fit the EDGES experimental bounds, the parameter is fixed at
$\beta_0=5.7_{-3.3}^{+3.9}\times10^{65}$.
\begin{figure}[H]
 \centering
  \includegraphics[scale=0.8]{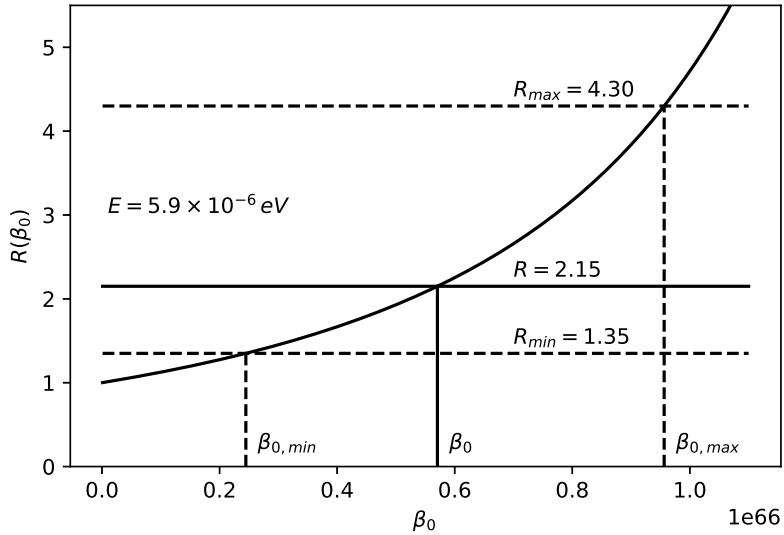}\\
  \caption{$R$ vs $\beta_0$ for fixed energy $E=E_{12}\simeq 5.9\times 10^{-6}$eV. }\label{Rbeta}
\end{figure}

\end{itemize}

\item Case 2: $f\left( E/{{E}_{P}} \right)=\frac{{{e}^{\alpha E/Ep}}-1}{\alpha E/{{E}_{P}}}$, $g\left( E/{{E}_{P}} \right)=1~$. The ratio $R$ reads
 \[
 R=\frac{e^{\alpha E/E_{P}} (e^{\alpha E/E_{P}} -1)^2}{(\alpha E/E_P)^2 }\,.
 \]
 The ratio $R$ is plotted as a function of $\alpha$ in Fig. \ref{REPL2}. To fit the EDGES experimental bounds, the parameter is fixed at
$\alpha=7.8_{-4.7}^{+6.9}\times10^{32}$.

\begin{figure}[H]
 \centering
  \includegraphics[scale=0.8]{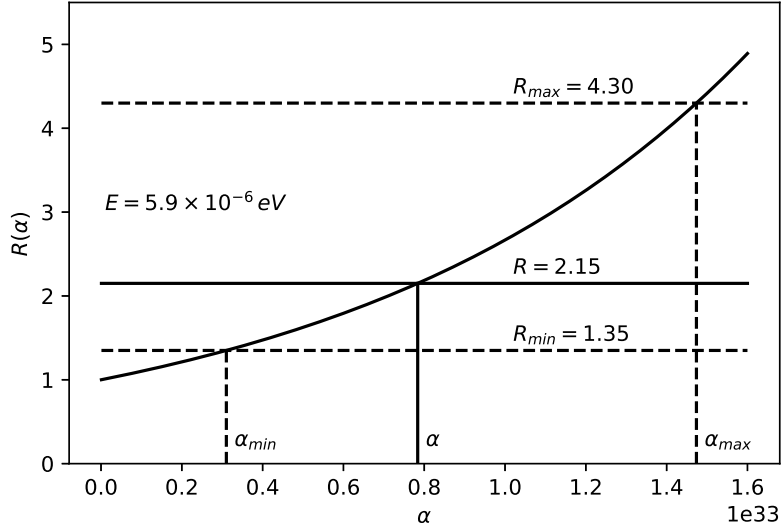}\\
  \caption{$R$ vs $\alpha$ for fixed energy $E=E_{12}\simeq 5.9\times 10^{-6}$eV. }\label{REPL2}
\end{figure}

\item Case 3: $f=1$, $g=\left[1+\lambda'\, (E/E_P)^\gamma\right]^\delta$. The ratio $R$ reads
\begin{eqnarray}
\label{Ral3}
R=\frac{|1+(1-\delta\gamma)\lambda'\, (E/E_P)^\gamma|}{[1+\lambda'\, (E/E_P)^\gamma]^{3\delta+1}}~,
\end{eqnarray}
for arbitrary parameters $\lambda'$, $\gamma$ and $\delta$. We take a look at the special case: 

\begin{itemize}
    \item[i)] For $\delta=1$, $\gamma=1$ and $\lambda'=\pm\alpha_0~$, we have \begin{eqnarray}
    R=\frac{1}{(1\pm\alpha_0E/E_P)^4}~.
    \end{eqnarray}
    The ratio $R$ is plotted as a function of $\alpha_0$ for both branches in Fig. \ref{Ralpha2}. However, only the branch with $\lambda'=-\alpha_0$ can fix $\alpha_0$. To fit the EDGES experimental bounds, the parameter is fixed at
$\alpha_0=3.6_{-2.1}^{+2.7}\times10^{32}$.
    \begin{figure}[H]
 \centering
  \includegraphics[scale=0.8]{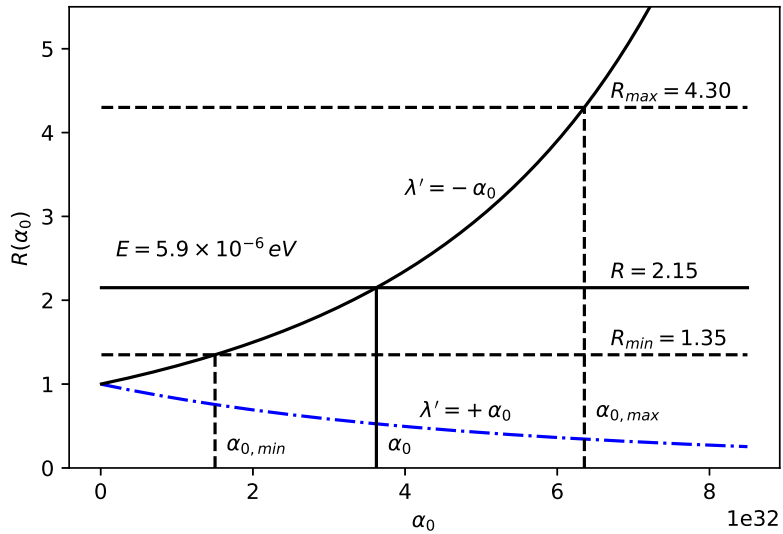}\\
  \caption{$R$ vs $\alpha_0$ for fixed energy $E=E_{12}\simeq 5.9\times 10^{-6}$eV. The $\lambda'=+\alpha_0$ branch is presented in dash-dot blue and the $\lambda=-\alpha_0$ branch is presented in solid black.}\label{Ralpha2}
\end{figure}
\end{itemize}
\end{itemize}
\par\noindent
At this point it should be stressed that the above plots indicate that the MDRs provided by cases 1, 2 and 3, give  $(\eta, \alpha_0, \sqrt{\beta_0}, \alpha, \lambda')\left|_{z=z_E} \approx 10^{32}\right.$ at redshift $z=z_E$.
These values are much larger than the bound set by the electroweak scale $\lambda_{EW}/\ell_P\lesssim 10^{17}$. 
To verify the compatibility with known observations and 
obtain the bounds on the above parameters in the current epoch ($z=0$), we compare the experimental precision of the CMB temperature $\left(\frac{\delta T}{T}\right)_{exp}=2\times10^{-4}$ \cite{Fixsen:2009ug} (see also Refs. \cite{Fixsen:1996nj,Mather:1998gm}) of a perfect black body to the theoretical deviation due to MDRs in the current epoch
\begin{eqnarray}
\label{relT}
\frac{\delta T}{T}(z=0)=(R(E)-1)\frac{\cosh{\left(\beta_{T_{CMB}}\!(0)\,E\right)}-1}{e^{\beta_{T_{CMB}}\!(0)\,E}-1}\,\frac{2}{\beta_{T_{CMB}}\!(0)\,E}~.
\end{eqnarray}
In the above, $R(E)$ is given by Eq. (\ref{R}) and $\beta_{T_{CMB}}\!(0)$ is given in terms of the CMB temperature in the current epoch. We obtain Eq. (\ref{relT}) by expressing $\frac{\delta T}{T}$ from $\rho_{MDR}(E,T)=\rho(E,T) R\approx\rho(E,T)+\frac{\mathrm{d}\rho}{\mathrm{d}T}(E,T)\,\delta T$. The parameters in the current epoch then must satisfy an upper bound of $(\eta, \alpha_0, \sqrt{\beta_0}, \alpha, \lambda')\left|_{z=0} < 10^{28}\right.$ to be consistent with the observed CMB spectrum in the current epoch. The bound obtained from the electroweak experiments is stronger than that, so it should be used as the relevant MDR parameter bound in the current epoch. The EDGES anomaly at $z=z_E$ combined with the 
above bound at $z=0$
suggest that the above parameters should be increasing functions of the redshift $z$. Therefore, we also expect $R$ to increase with $z$ for a given energy $E$ and have a value of $R\approx1$ at $z=0$. 

The compatibility of such MDRs with epochs earlier than $z_E$ should be taken into consideration as well. For example, in the epoch of the Big Bang Nucleosynthesis (BBN), at $z\approx3\times10^8$ \cite{BSR}, a bound of $\beta_0\lesssim10^{87}$ was obtained in \cite{Luciano:2021vkl} for the quadratic GUP parameter $\beta_0$, which corresponds to an upper bound $\lesssim10^{44}$ for the MDR parameters. Therefore, the values of the MDR parameters, measured by the EDGES anomaly are consistent with the BBN measurements, even if they increase to $\sim10^{44}$ at $z\approx3\times10^{8}$. This supports the increasing trend of the redshift dependence of the MDR parameters and may in fact provide a clue in determining the exact form of this dependence. Estimations of the MDR parameters from the modified CMB spectrum would not be relevant in the BBN epoch, since it has not been created until the epoch of recombination at $z=1090$ \cite{BSR}.

The standard MDRs used in this work can be found in Refs. \cite{Amelino-Camelia:2000stu,Amelino-Camelia:1996bln,Amelino-Camelia:2008aez,Majhi:2013koa,Das:2021lrb,Alexander:2001ck} as mentioned in Section \ref{mdr}, but they consider the MDR parameters as constants. The assumption that the MDR parameters are functions of another parameter, such as redshift, is fairly new. However, such an assumption is indirectly supported by Ref. \cite{Ong:2018zqn}, where the author finds a mass/radius dependent GUP parameter. This is also supported by the difference between estimations of the quadratic GUP parameter in tabletop experiments, where $\beta_0>0$ \cite{IPK,SLV,KP,Das:2021skl,Bosso:2016ycv}, and astrophysical/cosmological observations, where $\beta_0<0$ \cite{Das:2021nbq,Ong:2018zqn,Nenmeli:2021orl,JizbaScard,BuoninfCorp,JizbaLamb}. This shows that the MDR parameters can in fact be dependent on scale or redshift.

Since the usual models of modified dispersion relations can not explain the EDGES anomaly, it is also legitimate to investigate if it can be explained by considering  the cases analyzed here in which $\eta, \alpha_0, \lambda'=10^{17}$, namely they are fixed to the electroweak scale, while  $\omega$ and $\delta$ are treated as free parameters. We only consider cases 1 and 3, since case 2 has no other parameters to tweak. Also, we did not separately consider the special case 1, iii), because it is automatically considered as $\omega\longrightarrow2$.

\begin{figure}[H]
  \centering
  \includegraphics[scale=0.8]{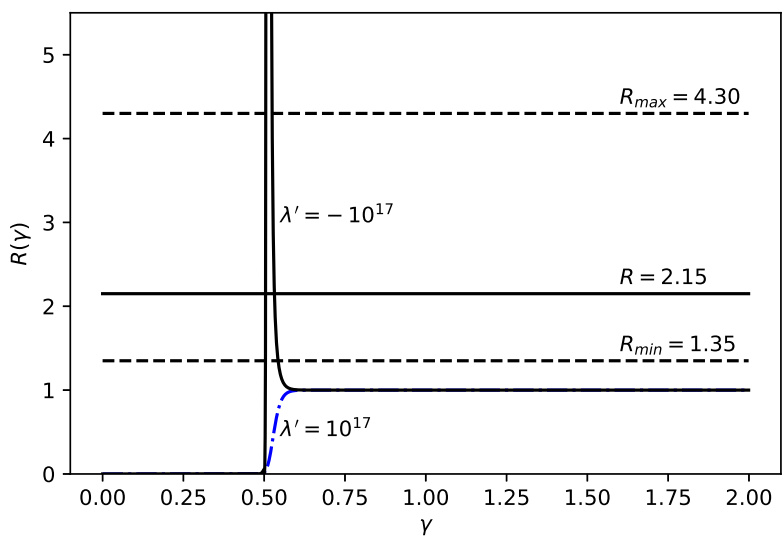}\\
  \caption{$R$ vs $\gamma$ for fixed energy $E=E_{12}\simeq 5.9\times 10^{-6}$eV. The $\lambda'=+10^{17}$ branch is presented in dash-dot blue and the $\lambda'=-10^{17}$ branch is presented in solid black.}\label{R3NC}
\end{figure}



\begin{figure}[H]
  \centering
  \includegraphics[scale=0.8]{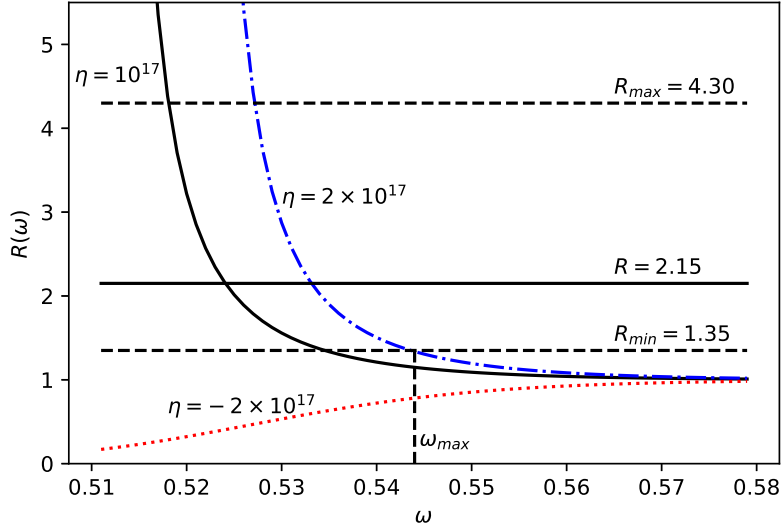}\\
  \caption{$R$ vs $\omega$ for fixed $\eta,\alpha_0=10^{17}$ and energy $E=E_{12}\simeq 5.9\times 10^{-6}$eV. The solid black, the dash-dot blue and dotted red lines represent cases 1) i) and ii) (positive and negative branch) respectively.}\label{R1Mod}
\end{figure}
\vspace{0.4cm}
\begin{figure}[H]
  \centering
  \includegraphics[scale=0.8]{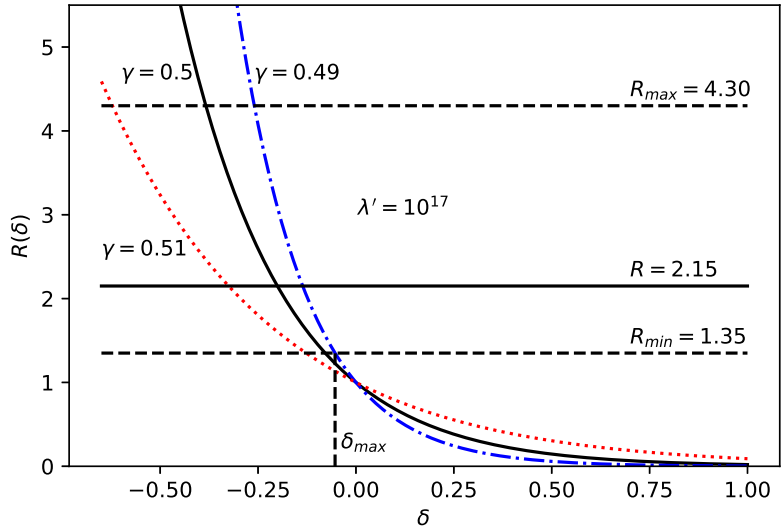}\\
  \caption{$R$ vs $\delta$ for fixed $\lambda'=10^{17}$ and energy $E=E_{12}\simeq 5.9\times 10^{-6}$eV. The dash-dot blue, solid black and dotted red lines represent $\gamma=0.49,\,0.50,\,0,51$ respectively.}\label{R3Mod}
\end{figure}
In Fig. \ref{R3NC}, we plot $R$ from Eq. (\ref{Ral3}) vs $\gamma$ for fixed $\lambda' =\pm10^{17}$ and for fixed $\delta=1$. The values of $R$ for $\lambda'=10^{17}$ fall outside the EDGES bounds and cannot provide an explanation for the EDGES anomaly. However, the values of $R$ for $\lambda'=-10^{17}$ fall inside the EDGES bounds twice in a narrow range of $\gamma$ around $\gamma\approx0.5$ and can therefore provide an explanation for the EDGES anomaly.
Changing the $\delta$ parameter only moves the peak to a different location.

%
%
At this point a number of comments are in order.
First, the power dependencies on $\omega$ and $\delta$ of these cases are shown in Figs. \ref{R1Mod} and \ref{R3Mod}, respectively. It is easily seen that the EDGES anomaly can be explained by powers $\omega_{max}< 0.544$ and $\delta_{max}< -0.05$. We also notice that we can only set an upper bound to the powers $\omega$ and $\delta$, since the electroweak length scale $\lambda_{EW}$ is an upper bound for the new length scale.
Second, the stringent values for $\omega$ and $\delta$, to resolve the EDGES anomaly, with their respective errors will be available in the future, when the true new length scale will be estimated and known with higher energy accelerators and astrophysical observations. 
Third, we also point out that power $\delta<0$, which means that the correction decreases with increasing $E$ as also seen in case 1. It may be noted that 
negative $\delta$ and positive $\lambda'$ is equivalent to positive $\delta$ and negative $\lambda'$ to leading order.
%
%
%
%
\section{Conclusion}
\label{conclusions}
%
%
%
%
\par\noindent
In this work, we have studied the possibility that MDRs can account for the recent results of the EDGES collaboration, which has discovered an anomalous absorption signal in the CMB radiation spectrum. This signal is larger by about a factor of $2$ with respect to the expected value (assuming that the background is described by the $\Lambda$CDM model), i.e., the EDGES anomaly.
In particular, we have shown that the 
most commonly considered MDRs, namely cases 1-3, lead to a modified thermal spectrum and to the
subsequent estimation of the parameters $(\eta, \alpha_0, \sqrt{\beta_0}, \alpha, \lambda')\left|_{z=z_E} \approx 10^{32}\right.$.  
Unfortunately, the parameter values at redshift $z=z_E$ are
outside the bounds allowed by, e.g., the electroweak experiments, since $(\eta, \alpha_0, \sqrt{\beta_0}, \alpha, \lambda')\left|_{z=z_E} \gg\lambda_{EW}/\ell_P=10^{17}\right.$. However, given the precision of the CMB temperature in the current epoch, $z=0$, we were able to constrain these parameters to an upper bound $(\eta, \alpha_0, \sqrt{\beta_0}, \alpha, \lambda')\left|_{z=0}\right. <10^{28}$ to be consistent with the observed CMB black body spectrum. The estimation of the MDR parameters from the EDGES anomaly at $z=z_E$, the bound obtained from electroweak experiments at $z=0$ and the BBN bound at $z\approx3\times10^8$  
suggest that the MDR parameters should be functions of redshift $z$ and as such could explain the EDGES anomaly. We can assume that the evolution of MDR parameters with time in the current epoch is slow or nearly constant, since we observe the same physics in all observable astrophysical objects such as distant galaxies. However, the time evolution of MDR parameters could have been faster in the early stages of the Universe as the EDGES anomaly suggests.

There is also another way out! 
As seen in Figs. \ref{R3NC}, \ref{R1Mod} and \ref{R3Mod} and explained there,
letting the powers $\omega$, $\gamma$ and $\delta$  
vary does also explain the anomaly for finite ranges of those powers. 
To precisely fit the EDGES experiment, and set $\eta, \alpha_0, \sqrt{\beta_0}, \alpha, \lambda'=10^{17}$, bounded by the electroweak scale, we have studied the possibility of treating the powers $\omega$, $\gamma$ and $\delta$ of the MDRs as free parameters and estimating upper bounds to their values. Similar results were found in Ref. \cite{Arzano:2016twc}. However, MDRs with non-trivial power dependencies require further research to better understand their importance for QG theories.


The results in this work indicate that MDRs originating from existing theories and thought experiments with minimal measurable length can provide a mechanism which explains the EDGES anomaly only if the MDR parameters are increasing functions of redshift $z$. Also, if the true QG theory with minimum measurable length predicts non-trivial deformation parameters as obtained from Figs. \ref{R3NC}, \ref{R1Mod} and \ref{R3Mod}, then such a theory can also provide a viable mechanism to explain the EDGES anomaly as well. It will be interesting to study the consequences of such deformation parameters in various contexts, such as GRBs physics \cite{Amelino-Camelia:1997ieq}. We hope to report on this in the future. 
%
%
%
%
\section{Acknowledgement}
%
%
%
%
%
\par\noindent
SD and ECV would like to acknowledge networking support by the COST Action CA18108.
This work was supported by the Natural Sciences and Engineering Research Council of Canada. 
%
%
%
%
\appendix
%
%
%
%
\section{21-cm cosmology}
\label{appendix}
%
%
%
%
\par\noindent
In this Appendix, we briefly recall the main features of the 21-cm cosmology. First, we note that the 21-cm line is associated with the relative orientation of electron and proton spins (anti-parallel for the singlet level with lower energy $E_{\uparrow\downarrow}$, parallel for the triplet level with higher energy $E_{\uparrow\uparrow}$). This gives rise to a hyperfine energy splitting between the two energy levels of the 1s ground state of  the hydrogen atom. 
The corresponding energy gap $E_{21}\equiv E_{\uparrow\uparrow}-E_{\uparrow\downarrow}$, hence that of the absorbed or emitted photons, 
is given by $E_{21} = 5.87\,\mu{\rm eV}$, which corresponds to a wavelength $\lambda_{21}^{\rm rest}=21$ cm or frequency
$\nu_{21}^{\rm rest} = 1420\,{\rm MHz}$. Due to this 21-cm transition, a neutral hydrogen at the recombination, with redshift $z \lesssim z_{\rm rec}$, can act as a detector of the background photons that have been produced at higher redshifts.
In the $\Lambda$CDM model such a photon background is given by the thermal radiation of the CMB with temperature
$T_{CMB}(z) = T_{CMB,0}\, (1+z)$, where $T_{CMB,0} = 2.725\,K \simeq 2.35 \times 10^{-4}\,{\rm eV}$.

The frequency of the 21-cm transition falls in the Rayleigh-Jeans tail, where the intensity $I_\nu\propto T$. To study absorption and emission of light, we can therefore use the integrated radiative transfer equation (in a rest frame) written in terms of temperature \cite{Pritchard:2011xb,Zaldarriaga:2003du}
\begin{eqnarray}
T_b(\tau_\nu)=T_S\left(1-e^{-\tau_\nu}\right)+T_{\gamma}\,\,e^{-\tau_\nu}~,
\end{eqnarray}
where $T_b(\nu)$ is the observed absolute brightness temperature, $T_S(z)$ the so-called {\em spin temperature} defined by the ratio of the population of the excited state $n_2$
with respect to the ground state states $n_1$,
\begin{eqnarray}
\label{boltzeq}
\frac{n_2}{n_1}(z) \equiv \frac{g_2}{g_1}  e^{-\frac{E_{21}}{k_BT_S(z)}}~,
\end{eqnarray}
(where $g_2/g_1 =3$ indicates the ratio of the statistical degeneracy factors of the two levels) and $\tau_\nu$ is the optical depth (of the hydrogen cloud in our case) defined as \cite{Furlanetto:2006jb,Pritchard:2011xb}
\begin{eqnarray}
\label{optdep}
\tau_\nu=\int \mathrm{d}s\, \sigma_{21}\left(1-e^{-\frac{E_{21}}{k_BT_S}}\right)\phi(\nu)\,n_0\approx \sigma_{21}\left(\frac{h\nu}{k_BT_S}\right)\left(\frac{N_{\mathrm{HI}}}{4}\right)\phi(\nu)~.
\end{eqnarray}
In the above, $\phi(\nu)$ is the line profile which, in general, is a Voigt function normalized as $\int\phi(\nu)\mathrm{d}\nu=1$, $\mathrm{d}s$ is the line element between the source and the observer, $n_0$ is the number density of neutral hydrogen, $N_{\mathrm{HI}}=\int \mathrm{d}s\,n_0$ is the column density of neutral hydrogen and $\sigma_{21}$ is the absorption cross-section for the transition. 
The latter is defined as \cite{Furlanetto:2006jb}
\begin{eqnarray}
\label{crosssec}
\sigma_{21}=\frac{3\,c^2A_{21}}{8\pi\nu^2}~,
\end{eqnarray}
where $A_{21}$ is the Einstein coefficient for spontaneous emission.
A relevant quantity in context of the 21-cm cosmology is the brightness temperature (the 21-cm brightness temperature is expressed relatively to the photon background at redshift $z$) \cite{Furlanetto:2006jb,Pritchard:2011xb,Zaldarriaga:2003du}
\begin{eqnarray}
\label{T21}
T_{21}(z) &\equiv& \delta T_b (z)=\frac{T_S(z)-T_{\gamma}(z)}{1+z}(1-e^{-\tau_\nu})\simeq \frac{T_S(z)-T_{\gamma}(z)}{1+z}\tau_{\nu} \nonumber \\
&\simeq&  23\,{\rm mK} \, (1+\delta_B)\, x_{H_I}(z) \,\left(\frac{\Omega_B\,h^2}{0.02}\right)\,
\left(\frac{0.15}{\Omega_{\rm m}\,h^2}\right)^{1/2}\,
\left(\frac{1+z}{10} \right)^{1/2}  \,\left[
1 - \frac{T_\gamma(z)}{T_S(z)} \right]\,,
\end{eqnarray}
where $\Omega_B\,h^2 = 0.02226$ is the baryon abundance, $\Omega_{\rm m}\,h^2=0.1415 $ matter abundances \cite{Planck:2015fie},
$\delta_B$ the baryon overdensity, $x_{H_I}$ the fraction of neutral hydrogen, and $T_{\gamma}(z)$ the effective temperature
(at frequency $\nu_{21}(z)(=\nu_{21}^{\rm rest}/(1+z))$) of the photon background radiation (in the $\Lambda$CDM model it does coincide with
$T_{CMB}(z)$).  
The spin temperature $T_S$ is related to the kinetic temperature of the gas $T_{\rm gas}$ by
\begin{equation}
1- \frac{T_{\gamma}}{T_S} \simeq \frac{x_c + x_\alpha}{1 + x_c + x_\alpha }\, \left(1- \frac{T_{\gamma}}{T_{\rm gas}}\right)\,.
\end{equation}
Here the coefficients $x_{c}$ and $x_{\alpha}$ describe the coupling between the hyperfine levels and the gas, characterized by the fact that
for $x_\alpha + x_c \gg 1$ (limit of strong coupling)
it follows $T_S = T_{\rm gas}$, while for $x_\alpha = x_c =0$ (no coupling), it follows $T_S = T_{\gamma}$, which means
that there is no signal.

%
%
%
%
\section{Modification of Einstein coefficients}
\label{appendix1}
%
%
%
%
Considering a gas of hydrogen, we study the MDR modification of absorption, spontaneous emission, and induced emission which take place when there is background radiation, with specific frequency passing through the gas. This will in turn provide a mechanism to study the MDR modification to the optical depth $\tau_\nu$. In this Appendix we follow the procedure outlined in \cite{Hilborn:1982}.

If there are $N_2$ atoms in a higher energy state, the atoms will spontaneously decay to a lower energy state and emit photons with a specific frequency $\nu=\Delta E/h$, where $\Delta E$ is the energy difference between the two states (in our case it will be energy $E_{21}$ of the hyperfine transition). The transition rate for spontaneous emission is 
\begin{eqnarray}
\label{sem}
W_{21}^s=A_{21}N_2~,
\end{eqnarray}
where $A_{21}$ is the Einstein coefficient for spontaneous emission. Note that here 1 and 2 refer to the lower and higher energy states respectively, and not as a subscript for 21-cm. 

On the other hand, if we have $N_1$ atoms in the lower energy state and subject them to radiation with thermal spectrum $\rho_{MDR}(\nu)=\rho(\nu)R$ and frequency $\nu$ (which is a frequency corresponding to the hyperfine energy splitting), the radiation will be absorbed, and the atoms will transition to the higher energy state. The transition rate for induced absorption is
\begin{eqnarray}
\label{iabs}
W_{12}=B_{12}N_1\rho(\nu)R~,
\end{eqnarray}
where $B_{12}$ is the Einstein coefficient for induced absorption and $R$ is the MDR modification of the thermal spectrum originating from Eq. (\ref{rhoNC}).

There is also a third possibility, where we have $N_2$ atoms in the higher energy state and subject them to radiation with thermal spectrum $\rho_{MDR}(\nu)$ and frequency $\nu$, the radiation will induce emission of new photons, originating from the transition to a lower level and traveling in the same direction as the incident radiation. The transition rate for induced emission is
\begin{eqnarray}
\label{iem}
W_{21}^i=B_{21}N_2\rho(\nu)R~,
\end{eqnarray}
where $B_{21}$ is the Einstein coefficient for induced emission. The total transition rate for emission is the sum of the spontaneous and induced emission transition rates
\begin{eqnarray}
W_{21}=W_{21}^s+W_{21}^i=N_2\left(A_{21}+B_{21}\rho(\nu)R\right)~.
\end{eqnarray}

The principle of detailed balance tells us that in thermal equilibrium the emission and absorption rates are equal, i.e., $W_{21}=W_{12}$. Using this, as well as 
Eqs. (\ref{boltzeq}), (\ref{radiation})
and $E=h\nu$, 
we obtain the ratio of the Einstein coefficients
\begin{eqnarray}
\label{ratioA}
\frac{A_{21}}{B_{21}}=\frac{8\pi h\nu^3}{c^3}R~,
\end{eqnarray}
and the ratio of induced absorption and emission coefficients
\begin{eqnarray}
\label{ratioB}
\frac{B_{12}}{B_{21}}=\frac{g_2}{g_1}~.
\end{eqnarray}
From Eq. (\ref{ratioA}) we see that while the ratio of the Einstein coefficients is modified, it does not tell us how the individual coefficients are modified. 
However, Eqs. (\ref{sem}), (\ref{iabs}) and (\ref{iem}) suggest that only the $B$ coefficients are modified by $R$ and the $A$ coefficient remains unmodified, since
\begin{eqnarray}
B_{12}\propto B_{21}\propto \frac{1}{R}\,\,\,\,\,\,\,\,\,\,\mathrm{and}\,\,\,\,\,\,\,\,\,\,A_{21}\not\propto f(R)~.
\end{eqnarray}

Next we show that the 
absorption cross-section $\sigma_{21}$ does not depend on $R$ as well. 
%
%
The driving equation to study absorption and emission of light in a gas is the radiative transfer equation, written in the differential form
\begin{eqnarray}
\label{radtrans}
\frac{\mathrm{d}I_\nu}{\mathrm{d}s}=-\,\alpha(\nu)\,I_{\nu}~,
\end{eqnarray}
where we neglect the emission part (we only need information on absorption, since we are looking for the absorption cross-section), $I_\nu=c\rho(\nu)R=I_{0\nu}R$ is the spectral intensity, $I_{0\nu}$ is the unmodified spectral intensity and $\alpha(\nu)$ is the absorption coefficient, related to the absorption cross-section as
\begin{eqnarray}
\label{abscoeff}
\alpha(\nu)=n_1\,\sigma_{21}\,\phi(\nu)~.
\end{eqnarray}
By plugging $I_\nu=I_{0\nu}R$ in Eq. (\ref{radtrans}), we find that the same radiative transfer equation holds also for $I_{0\nu}$, since $R\neq f(s)$ and reads as
\begin{eqnarray}
\label{radtrans0}
\frac{\mathrm{d}I_{0\nu}}{\mathrm{d}s}=-\,\alpha(\nu)\,I_{0\nu}~.
\end{eqnarray}

The power of the incident beam with frequencies between $\nu$ and $\nu+\mathrm{d}\nu$ is absorbed by $N_1$ atoms reads as
\begin{eqnarray}
-\Delta P=h\nu\, W_{12}\,\phi(\nu)\mathrm{d}\nu=h\nu B_{12}N_1\rho(\nu)\,R\,\phi(\nu)\mathrm{d}\nu~,
\end{eqnarray}
where $h\nu$ is the energy of a photon with frequency $\nu$, $W_{12}$ is the absorption transition rate defined in Eq. (\ref{iabs}) and $\phi(\nu)$ is the line profile defined in Eq. (\ref{optdep}). Writing the number of atoms in the ground state as $N_1=n_1A\Delta s$, where the atoms are confined in a volume $A\Delta s$ and the thermal spectrum $\rho(\nu)$ in terms of spectral intensity $I_\nu$, we get
\begin{eqnarray}
\label{delpow}
-\Delta P=\frac{h\nu}{c}B_{12}\,n_1A\,\Delta s\,I_{0\nu} R\,\phi(\nu)\mathrm{d}\nu~.
\end{eqnarray}
By the definition of the spectral intensity, we know that $\frac{\Delta P}{A\mathrm{d}\nu\Delta s} \implies \frac{\mathrm{d}I_{0\nu}}{\mathrm{d}s}$. Therefore we can rewrite Eq. (\ref{delpow}) as
\begin{eqnarray}
\frac{\mathrm{d}I_{0\nu}}{\mathrm{d}s}=-\frac{h\nu}{c}n_1B_{12}\,R\,\phi(\nu)\,I_{0\nu}~.
\end{eqnarray}
By comparing the above equation with Eqs. (\ref{radtrans0}) and (\ref{abscoeff}), we obtain the absorption cross-section
\begin{eqnarray}
\sigma_{21}=\frac{h\nu}{c}B_{12}\,R~.
\end{eqnarray}
By using Eqs. (\ref{ratioB}) and (\ref{ratioA}) in the above, we see that the factor $R$ cancels and we obtain the final expression for the absorption cross-section
\begin{eqnarray}
\sigma_{21}=\frac{3\,c^2A_{21}}{8\pi\nu^2}~,
\end{eqnarray}
which is exactly the same as Eq. (\ref{crosssec}). We see that the above absorption cross-section does not depend on $R$ and therefore, according to Appendix \ref{appendix}, MDRs do not modify the optical depth $\tau_\nu$.


\begin{thebibliography}{99}

\bibitem{Banks:1996vh}
T.~Banks, W.~Fischler, S.~H.~Shenker and L.~Susskind,
Phys. Rev. D \textbf{55}, 5112-5128 (1997)
[arXiv:hep-th/9610043 [hep-th]].


\bibitem{Seiberg:1999vs}
N.~Seiberg and E.~Witten,
JHEP \textbf{09}, 032 (1999)
[arXiv:hep-th/9908142 [hep-th]].


\bibitem{Sheikh-Jabbari:1999cvv}
M.~M.~Sheikh-Jabbari,
Phys. Lett. B \textbf{455}, 129-134 (1999)
[arXiv:hep-th/9901080 [hep-th]].


\bibitem{Maldacena:2000vw}
J.~M.~Maldacena and J.~G.~Russo,
Class. Quant. Grav. \textbf{17}, 1189-1203 (2000).


\bibitem{Chamseddine:2001hk}
A.~H.~Chamseddine and M.~S.~Volkov,
JHEP \textbf{04}, 023 (2001)
[arXiv:hep-th/0101202 [hep-th]].


\bibitem{Hashimoto:2000ys}
A.~Hashimoto and N.~Itzhaki,
Phys. Rev. D \textbf{63}, 126004 (2001)
[arXiv:hep-th/0012093 [hep-th]].


\bibitem{Alexander:2001dr}
S.~Alexander, R.~Brandenberger and J.~Magueijo,
Phys. Rev. D \textbf{67}, 081301 (2003)
[arXiv:hep-th/0108190 [hep-th]].


\bibitem{Kempf:2000ac}
A.~Kempf,
Phys. Rev. D \textbf{63}, 083514 (2001)
[arXiv:astro-ph/0009209 [astro-ph]].


\bibitem{Capozziello:1999wx}
S.~Capozziello, G.~Lambiase and G.~Scarpetta,
Int. J. Theor. Phys. \textbf{39}, 15-22 (2000)
[arXiv:gr-qc/9910017 [gr-qc]].


\bibitem{Chu:2000ww}
C.~S.~Chu, B.~R.~Greene and G.~Shiu,
Mod. Phys. Lett. A \textbf{16}, 2231-2240 (2001)
[arXiv:hep-th/0011241 [hep-th]].


\bibitem{Easther:2001fi}
R.~Easther, B.~R.~Greene, W.~H.~Kinney and G.~Shiu,
Phys. Rev. D \textbf{64}, 103502 (2001)
[arXiv:hep-th/0104102 [hep-th]].


\bibitem{Kempf:2001fa}
A.~Kempf and J.~C.~Niemeyer,
Phys. Rev. D \textbf{64}, 103501 (2001)
[arXiv:astro-ph/0103225 [astro-ph]].

\bibitem{luciano} G.G. Luciano, L. Petruzziello, Eur. Phys. J. Plus {\bf 136} (2021). 

\bibitem{luciano1}
M. Blasone, G. Lambiase, G.G.Luciano, L. Petruzziello, L. Smaldone, Class. Quant. Grav. {\bf 37}, 155004 (2020).

\bibitem{Brandenberger:2000wr}
R.~H.~Brandenberger and J.~Martin,
Mod. Phys. Lett. A \textbf{16}, 999-1006 (2001)
[arXiv:astro-ph/0005432 [astro-ph]].


\bibitem{Martin:2000xs}
J.~Martin and R.~H.~Brandenberger,
Phys. Rev. D \textbf{63}, 123501 (2001)
[arXiv:hep-th/0005209 [hep-th]].


\bibitem{Das:2008kaa}
S.~Das and E.~C.~Vagenas,
Phys. Rev. Lett. \textbf{101}, 221301 (2008)
[arXiv:0810.5333 [hep-th]].


\bibitem{Ali:2011fa}
A.~F.~Ali, S.~Das and E.~C.~Vagenas,
Phys. Rev. D \textbf{84}, 044013 (2011)
[arXiv:1107.3164 [hep-th]].


\bibitem{Amelino-Camelia:1997ieq}
G.~Amelino-Camelia, J.~R.~Ellis, N.~E.~Mavromatos, D.~V.~Nanopoulos and S.~Sarkar,
Nature \textbf{393}, 763-765 (1998)
[arXiv:astro-ph/9712103 [astro-ph]].


\bibitem{Amelino-Camelia:1999jfz}
G.~Amelino-Camelia and S.~Majid,
Int. J. Mod. Phys. A \textbf{15}, 4301-4324 (2000)
[arXiv:hep-th/9907110 [hep-th]].


\bibitem{Amelino-Camelia:2000stu}
G.~Amelino-Camelia,
Int. J. Mod. Phys. D \textbf{11}, 35-60 (2002)
[arXiv:gr-qc/0012051 [gr-qc]].


\bibitem{Gambini:1998it}
R.~Gambini and J.~Pullin,
Phys. Rev. D \textbf{59}, 124021 (1999)
[arXiv:gr-qc/9809038 [gr-qc]].


\bibitem{tHooft:1996ziz}
G.~'t Hooft,
Class. Quant. Grav. \textbf{13}, 1023-1040 (1996)
[arXiv:gr-qc/9601014 [gr-qc]].


\bibitem{Kostelecky:1988zi}
V.~A.~Kostelecky and S.~Samuel,
Phys. Rev. D \textbf{39}, 683 (1989).


\bibitem{Carroll:2001ws}
S.~M.~Carroll, J.~A.~Harvey, V.~A.~Kostelecky, C.~D.~Lane and T.~Okamoto,
Phys. Rev. Lett. \textbf{87}, 141601 (2001)
[arXiv:hep-th/0105082 [hep-th]].


\bibitem{Horava:2009uw}
P.~Horava,
Phys. Rev. D \textbf{79}, 084008 (2009)
[arXiv:0901.3775 [hep-th]].


\bibitem{Horava:2009if}
P.~Horava,
Phys. Rev. Lett. \textbf{102}, 161301 (2009)
[arXiv:0902.3657 [hep-th]].



\bibitem{Amelino-Camelia:2000cpa}
G.~Amelino-Camelia,
Phys. Lett. B \textbf{510}, 255-263 (2001)
[arXiv:hep-th/0012238 [hep-th]].


\bibitem{Amelino-Camelia:2010lsq}
G.~Amelino-Camelia,
Symmetry \textbf{2}, 230-271 (2010)
[arXiv:1003.3942 [gr-qc]].






\bibitem{Magueijo:2002xx}
J.~Magueijo and L.~Smolin,
Class. Quant. Grav. \textbf{21}, 1725-1736 (2004)
[arXiv:gr-qc/0305055 [gr-qc]].




\bibitem{Ali:2014aba}
A.~F.~Ali and M.~M.~Khalil,
EPL \textbf{110}, no.2, 20009 (2015)
[arXiv:1408.5843 [gr-qc]].


\bibitem{Liu:2007fk}
C.~Z.~Liu and J.~Y.~Zhu,
Gen. Rel. Grav. \textbf{40}, 1899-1911 (2008)
[arXiv:gr-qc/0703055 [gr-qc]].


\bibitem{Leiva:2008fd}
C.~Leiva, J.~Saavedra and J.~Villanueva,
Mod. Phys. Lett. A \textbf{24}, 1443-1451 (2009)
[arXiv:0808.2601 [gr-qc]].


\bibitem{Li:2008gs}
H.~Li, Y.~Ling and X.~Han,
Class. Quant. Grav. \textbf{26}, 065004 (2009)
[arXiv:0809.4819 [gr-qc]].


\bibitem{Garattini:2011hy}
R.~Garattini and G.~Mandanici,
Phys. Rev. D \textbf{85}, 023507 (2012)
[arXiv:1109.6563 [gr-qc]].


\bibitem{Awad:2013nxa}
A.~Awad, A.~F.~Ali and B.~Majumder,
JCAP \textbf{10}, 052 (2013)
[arXiv:1308.4343 [gr-qc]].


\bibitem{Barrow:2013gia}
J.~D.~Barrow and J.~Magueijo,
Phys. Rev. D \textbf{88}, no.10, 103525 (2013)
[arXiv:1310.2072 [astro-ph.CO]].


\bibitem{Ali:2014xqa}
A.~F.~Ali,
Phys. Rev. D \textbf{89}, no.10, 104040 (2014)
[arXiv:1402.5320 [hep-th]].


\bibitem{Ali:2014cpa}
A.~F.~Ali, M.~Faizal and B.~Majumder,
EPL \textbf{109}, no.2, 20001 (2015)
[arXiv:1406.1980 [gr-qc]].

\bibitem{Bowman:2018yin}
J.~D.~Bowman, A.~E.~E.~Rogers, R.~A.~Monsalve, T.~J.~Mozdzen and N.~Mahesh,
Nature \textbf{555}, no.7694, 67-70 (2018)
[arXiv:1810.05912 [astro-ph.CO]].


\bibitem{Arzano:2016twc}
M.~Arzano and G.~Calcagni,
Phys. Rev. D \textbf{93}, no.12, 124065 (2016)
[arXiv:1604.00541 [gr-qc]].


\bibitem{Amelino-Camelia:1996bln}
G.~Amelino-Camelia, J.~R.~Ellis, N.~E.~Mavromatos and D.~V.~Nanopoulos,
Int. J. Mod. Phys. A \textbf{12}, 607-624 (1997)
[arXiv:hep-th/9605211 [hep-th]].



\bibitem{Amelino-Camelia:2008aez}
G.~Amelino-Camelia,
Living Rev. Rel. \textbf{16}, 5 (2013)
[arXiv:0806.0339 [gr-qc]].


\bibitem{Majhi:2013koa}
B.~R.~Majhi and E.~C.~Vagenas,
Phys. Lett. B \textbf{725}, 477-480 (2013)
[arXiv:1307.4195 [gr-qc]].


\bibitem{Das:2021lrb}
A.~Das, S.~Das, N.~R.~Mansour and E.~C.~Vagenas,
Phys. Lett. B \textbf{819}, 136429 (2021)
[arXiv:2101.03746 [gr-qc]].


\bibitem{Alexander:2001ck}
S.~Alexander and J.~Magueijo,
``Noncommutative geometry as a realization of varying speed of light cosmology,''
[arXiv:hep-th/0104093 [hep-th]].




\bibitem{caniou}
J. Caniou, \emph{Passive Infrared Detection: Theory and Applications}, Boston, MA, USA: Springer-Verlag US (1999).

\bibitem{sharkov}
E. A. Sharkov, \emph{Passive Microwave Remote Sensing of the Earth}, Berlin, DE: Springer-Verlag Berlin Heidelberg (2003).



\bibitem{Barkana:2018lgd}
R.~Barkana,
Nature \textbf{555}, no.7694, 71-74 (2018)
[arXiv:1803.06698 [astro-ph.CO]].


\bibitem{Furlanetto:2006jb}
S.~Furlanetto, S.~P.~Oh and F.~Briggs,
Phys. Rept. \textbf{433}, 181-301 (2006)
[arXiv:astro-ph/0608032 [astro-ph]].


\bibitem{Fornengo:2011cn}
N.~Fornengo, R.~Lineros, M.~Regis and M.~Taoso,
Phys. Rev. Lett. \textbf{107}, 271302 (2011)
[arXiv:1108.0569 [hep-ph]].


\bibitem{Fixsen:2009xn}
D.~J.~Fixsen, \textit{et al.}
Astrophys. J. \textbf{734}, 5 (2011)
[arXiv:0901.0555 [astro-ph.CO]].

  
\bibitem{Pritchard:2011xb}
J.~R.~Pritchard and A.~Loeb,
Rept. Prog. Phys. \textbf{75}, 086901 (2012)
[arXiv:1109.6012 [astro-ph.CO]].


\bibitem{Lopez-Honorez:2016sur}
L.~Lopez-Honorez, O.~Mena, \'A.~Molin\'e, S.~Palomares-Ruiz and A.~C.~Vincent,
JCAP \textbf{08}, 004 (2016)
[arXiv:1603.06795 [astro-ph.CO]].


\bibitem{AristizabalSierra:2018emu}
D.~Aristizabal Sierra and C.~S.~Fong,
Phys. Lett. B \textbf{784}, 130-136 (2018)
[arXiv:1805.02685 [hep-ph]].

  
\bibitem{Hill:2018lfx}
J.~C.~Hill and E.~J.~Baxter,
JCAP \textbf{08}, 037 (2018)
[arXiv:1803.07555 [astro-ph.CO]].


\bibitem{Pospelov:2018kdh}
M.~Pospelov, J.~Pradler, J.~T.~Ruderman and A.~Urbano,
Phys. Rev. Lett. \textbf{121}, no.3, 031103 (2018)
[arXiv:1803.07048 [hep-ph]].

  
\bibitem{Moroi:2018vci}
T.~Moroi, K.~Nakayama and Y.~Tang,
Phys. Lett. B \textbf{783}, 301-305 (2018)
[arXiv:1804.10378 [hep-ph]].


\bibitem{Lambiase:2018lhs}
G.~Lambiase and S.~Mohanty,
Mon. Not. Roy. Astron. Soc. \textbf{494}, no.4, 5961-5966 (2020)
[arXiv:1804.05318 [hep-ph]].



\bibitem{Chianese:2018luo}
M.~Chianese, P.~Di Bari, K.~Farrag and R.~Samanta,
Phys. Lett. B \textbf{790}, 64-70 (2019)
[arXiv:1805.11717 [hep-ph]].

\bibitem{Fixsen:2009ug}
D.~J.~Fixsen,
Astrophys. J. \textbf{707}, 916-920 (2009)
[arXiv:0911.1955 [astro-ph.CO]].


\bibitem{Fixsen:1996nj}
D.~J.~Fixsen, E.~S.~Cheng, J.~M.~Gales, J.~C.~Mather, R.~A.~Shafer and E.~L.~Wright,
Astrophys. J. \textbf{473}, 576 (1996)
[arXiv:astro-ph/9605054 [astro-ph]].


\bibitem{Mather:1998gm}
J.~C.~Mather, D.~J.~Fixsen, R.~A.~Shafer, C.~Mosier and D.~T.~Wilkinson,
Astrophys. J. \textbf{512}, 511-520 (1999)
[arXiv:astro-ph/9810373 [astro-ph]].


\bibitem{BSR}
B. Ryden, \emph{Introduction to Cosmology 2nd Edition}, New York, USA: Cambridge University Press (2017).

\bibitem{Luciano:2021vkl}
G.~G.~Luciano,
Eur. Phys. J. C \textbf{81}, no.12, 1086 (2021)
[arXiv:2111.06000 [astro-ph.CO]].


\bibitem{Ong:2018zqn}
Y.~C.~Ong,
JCAP \textbf{09}, 015 (2018)
[arXiv:1804.05176 [gr-qc]].

\bibitem{IPK}
I.~Pikovski, M.~R.~Vanner, M.~Aspelmeyer, M.~S.~Kim and C.~Brukner,
Nature Phys. \textbf{8}, 393-397 (2012)
[arXiv:1111.1979 [quant-ph]].

\bibitem{SLV}
F. Scardigli, G. Lambiase and E. C. Vagenas, Phys. Lett. B \textbf{767}, 242-246 (2017),
[arXiv:1611.01469].

\bibitem{KP}
S.~P.~Kumar and M.~B.~Plenio,
Phys. Rev. A \textbf{97}, no.6, 063855 (2018)
[arXiv:1708.05659 [quant-ph]].

\bibitem{Bosso:2016ycv}
P.~Bosso, S.~Das, I.~Pikovski and M.~R.~Vanner,
Phys. Rev. A \textbf{96} (2017)  023849
[arXiv:1610.06796 [gr-qc]].

\bibitem{Das:2021skl}
S.~Das and M.~Fridman,
Phys. Rev. D \textbf{104}, 026014 (2021)
[arXiv:2104.04634 [gr-qc]].

\bibitem{Das:2021nbq}
S.~Das, M.~Fridman, G.~Lambiase and E.~C.~Vagenas,
Phys. Lett. B \textbf{824}, 136841 (2022)
[arXiv:2107.02077 [gr-qc]].


\bibitem{Nenmeli:2021orl}
V.~Nenmeli, S.~Shankaranarayanan, V.~Todorinov and S.~Das,
Phys. Lett. B \textbf{821}, 136621 (2021)
[arXiv:2106.04141 [gr-qc]].

\bibitem{JizbaScard}
P.~Jizba, H.~Kleinert and F.~Scardigli,
Phys. Rev. D \textbf{81}, 084030 (2010)
[arXiv:0912.2253 [hep-th]].

\bibitem{BuoninfCorp}
L.~Buoninfante, G.~G.~Luciano and L.~Petruzziello,
Eur. Phys. J. C \textbf{79}, 663 (2019) [arXiv:1903.01382 [gr-qc]].

\bibitem{JizbaLamb}
P.~Jizba, G.~Lambiase, G.~G.~Luciano and L.~Petruzziello,
Phys. Rev. D \textbf{105}, no.12, L121501 (2022)
[arXiv:2201.07919 [hep-th]].

  
\bibitem{Zaldarriaga:2003du}
M.~Zaldarriaga, S.~R.~Furlanetto and L.~Hernquist,
Astrophys. J. \textbf{608}, 622-635 (2004)
[arXiv:astro-ph/0311514 [astro-ph]].

  
\bibitem{Planck:2015fie}
P.~A.~R.~Ade \textit{et al.} [Planck Collaboration],
Astron. Astrophys. \textbf{594}, A13 (2016)
[arXiv:1502.01589 [astro-ph.CO]].

\bibitem{Hilborn:1982}
R.~C.~Hilborn, 
Am. J. Phys. \textbf{50}, 982–986 (1982)
[arXiv:physics/0202029 [physics.atom-ph]].


\end{thebibliography}
\end{document}